\documentclass[a4paper,11pt]{article}
\usepackage{jheppub} 
\usepackage[smartEllipses]{markdown}
\usepackage{ifthen}
\usepackage{easyReview}

\newcommand{\M}{\mathcal{M}}

\newcommand{\diag}{\operatorname{diag}}

\title{\boldmath Monodromy of multiloop integrals in $d$ dimensions}

\author{Roman N. Lee}
\author{Andrei A. Pomeransky}

\affiliation{Budker Institute of Nuclear Physics, Novosibirsk 630090, Russia}

\emailAdd{r.n.lee@inp.nsk.su}
\emailAdd{a.a.pomeransky@inp.nsk.su}
\allowdisplaybreaks
\abstract{
	We consider the monodromy group of the differential systems for multiloop integrals. We describe a simple heuristic method to obtain the monodromy matrices as functions of space-time dimension $d$. We observe that in a special basis the elements of these matrices are Laurent polynomials in $z=\exp(i\pi d)$ with integer coefficients, i.e., the monodromy group is a subgroup of $GL(n,\mathbb{Z}[z,1/z])$.
	We derive bilinear relations for monodromies in $d$ and $-d$ dimensions which follow from the twisted Riemann bilinear relations and check that the found monodromy matrices satisfy them.
}

\begin{document}

\maketitle
\flushbottom
\section{Introduction}
\label{sec:intro}

Multiloop integrals are the main building blocks in any perturbative calculation. As analytic functions of kinematic parameters, they exhibit a complicated branching structure. This structure is associated with the monodromy group of the integrals. The investigation of the monodromy group for the multiloop integrals has a long history, notably, including the work of Pham, Regge, Speer and collaborators in late 1960s \cite{pham1967singularites,regge1968algebraic,speer1969generalized,ponzano1969monodromy,ponzano1970monodromy}.

Starting from the 1980s, there has been a huge progress in multiloop calculations related to the development of the integration by parts (IBP)  method \cite{Tkachov1981,ChetTka1981}. This method provided an algorithmic approach to the derivation of the differential equations for dimensionally regularized multiloop integrals \cite{Remiddi1997,Kotikov1991}. The space-time dimension $d$ enters these equations as a parameter. These differential systems carry almost all essential information about integrals. In particular, their monodromy group is also fully determined by these systems. In fact, the close relation between the differential equations and monodromy group was emphasized already in Ref.  \cite{regge1968algebraic} (for a review, see Refs. \cite{golubeva1976some,golubeva2014regge}). In some sense, not only the differential system determines the monodromy group, but also the monodromy group determines the differential system up to a linear transformation of functions with rational coefficients.

In this paper, we introduce a simple and efficient method for calculating the monodromy of differential systems for multiloop integrals in $d$ dimensions. The method is based on some ansatz for the monodromy matrices as function of $d$ and uses the numerical recognition via integer relations algorithms to recover the analytic form.

We apply our method to several families of multiloop integrals. For each case, we derive explicit expressions for the matrices of monodromy group generators. Remarkably, in all considered examples we observe that all elements of monodromy group can be represented by matrices which form a subgroup of $GL(n,\mathbb{Z}[z,1/z])$, where $z=\exp(i\pi d)$, i.e., the matrices which elements are Laurent polynomials of $z$ with integer coefficients (their determinants should be equal to $\pm z^k$ with $k\in \mathbb{Z}$). In particular, in this representation the monodromy group is explicitly invariant under the shift $d\to d+2$. We show that twisted Riemann bilinear relations lead to the corresponding relations between the elements $\rho(\gamma,d)$ and  $\rho(\gamma,-d)$ of the monodromy representations in $d$ and $-d$ dimensions and check that the found monodromy representations satisfy them.

The paper is organized as follows.
In Section \ref{sec:monodromy} we recall some important facts and definitions related to the monodromy of linear differential systems, which can be found, e.g., in recent textbook \cite{haraoka2020linear}. Section \ref{sec:frobenius} explains the relation between the monodromy generators and generalized series expansion in the vicinity of singular points. In Section \ref{sec:d} we discuss the specific properties of the monodromy groups which appear when considering the multiloop integrals in $d$ dimensions. In Section \ref{sec:algorithm} we describe our approach to finding the monodromy as function of $d$. Section \ref{sec:examples} contains a number of examples of the application of our approach to various families of multiloop integrals in $d$ dimensions. Finally, in Section \ref{sec:conclusion}, we summarize our findings.

\section{Monodromy of linear differential system}\label{sec:monodromy}

Suppose that we have the differential system
\begin{equation}
	\partial_x J = M(x) J, \label{eq:de}
\end{equation}
where $M(x)$ is an $n\times n$ matrix  which elements are rational functions of a complex variable $x$ and $J=J(x)$ is a column of functions.

If we allow for M\"obius transformations $x\to \frac{Ax+B}{Cx+D}$, we can consider the system \eqref{eq:de} as defined on  $\overline{\mathbb{C}}\backslash S=\mathbb{CP}^1\backslash S$, where
\begin{equation}
	S=\{a_0,a_1,\ldots, a_p\}
\end{equation}
is the set of all singular points of the system. A finite point $x = a$ is called a singular point of this system   if $x = a$ is a pole of $M(x)$. Besides, we say that $\infty\in \overline{\mathbb{C}}$ is a singular point unless $\lim_{x\to\infty} xM(x) =0$. In what follows, we will assume that all singular points of the system are regular, i.e., that the general solution is bounded by a power function in the sectorial neighborhood of any singular point (see Ref. \cite[section 4.1]{haraoka2020linear}). In particular, a system of \emph{Fuchsian type}, i.e., of the form \eqref{eq:de} with
\begin{equation}
M(x)=\sum_{a\in S\backslash{\infty}} \frac{M_a}{x-a},\label{eq:fuchsian}
\end{equation}
where $M_a$ are constant matrices, has only regular singular points (the point $x=\infty$ is singular if $M_{\infty}=-\sum_aM_a$ is not zero).

\begin{figure}
	\centering
	\includegraphics[width=0.5\linewidth]{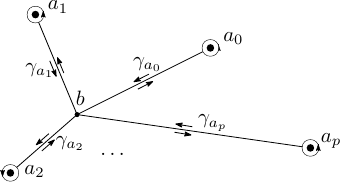}
	\caption{Generators of the fundamental group $\pi_1(\overline{\mathbb{C}}\backslash \{a_0,a_1,\ldots, a_p\},b)$. Each contour $\gamma_i$ goes from the base point $b$ along the ray in the direction of $a_i$, encircles $a_i$ and returns back. The points are numbered in anti-clockwise direction.}
	\label{fig:monodromy}
\end{figure}

In relation to the differential system \eqref{eq:de} it is natural to consider the fundamental group
\begin{equation}
G=\pi_1(\overline{\mathbb{C}}\backslash S,b),\label{eq:fundamental}
\end{equation}
which is generated by the loops $\gamma_{a_i}$ starting and ending at $x=b$, each encircling the corresponding singular point $a_i$ ($i=0,1,\ldots, p$), with a single relation. With an appropriate choice of these loops, the relation reads
\begin{equation}
	\gamma_{a_0}\gamma_{a_1}\ldots\gamma_{a_p}=1.
\end{equation}
In particular, for sufficiently generic $b$ we can choose the contours $\gamma_{a_i}$ as shown in Fig. \ref{fig:monodromy}.

A fundamental matrix $F(x)$ of the system \eqref{eq:de} determined in a neighborhood of $b$ can be analytically continued along any closed path $\gamma \in G$. The new fundamental matrix $F_\gamma(x)$ satisfies
\begin{equation}
	F_\gamma(x) =F(x)\rho(\gamma),
\end{equation}
where $\rho(\gamma) \in GL(n,\mathbb{C})$ is called the monodromy along $\gamma$. The monodromy representation $\rho:G\to GL(n,\mathbb{C})$ defines the monodromy group $\M=\rho(G)$.

It is important that not only the differential system determines the monodromy representation, but conversely, any $GL(n,\mathbb{C})$-representation of the fundamental group \eqref{eq:fundamental} can be associated with the monodromy representation of some differential system with regular singularities in $S$. This fact is closely related to Hilbert's 21st problem which asks if there exists such a differential system of Fuchsian type. In this formulation the problem has a negative solution, as shown by Bolibrukh \cite{bolibrukh1990riemann}. Nevertheless, the existence of a system with a given monodromy representation with regular singular points in $S$ has been proved already in 1908 by Plemelj \cite{plemelj1908riemannian} (see \cite[\S 1]{bolibrukh1990riemann}).

In fact, the representation of the fundamental group entirely determines the differential system up to the transformations
\begin{equation}
	J\to T(x)J,\label{eq:transformation}
\end{equation}
where $T(x)$ is an $n\times n$ matrix  which elements are rational functions of $x$, such that  $\det T(x)$ is not identically zero. Indeed, let $F_1(x)$ and $F_2(x)$ be the fundamental matrices for two differential systems with the same monodromy representation. Then the matrix elements of $T(x)=F_2(x)F_1^{-1}(x)$ are single-valued\footnote{Indeed, $T_\gamma(x)=[F_2(x)\rho(\gamma)] [F_1(x)\rho(\gamma)]^{-1}=F_2(x)F_1^{-1}(x)=T(x)$.} analytic functions on $\overline{\mathbb{C}}\backslash S$ bounded by power functions in sectorial neighborhoods of all points in $S$. Therefore, they are rational functions of $x$. Then $F_2(x)=T(x)F_1(x)$.

\section{Monodromy via generalized series expansions}\label{sec:frobenius}

Let $x=a$ be a regular singular point. We say that the system \eqref{eq:de} is in \emph{normalized Fuchsian form} at $a$ if the expansion of $M(x)$ at $x=a$ has the form
\begin{equation}
	M(x)=\frac{M_a}{x-a}+O\left(1\right),
\end{equation}
and the \emph{matrix residue} $M_a$ has no \emph{resonances}, i.e. eigenvalues which differ by a non-zero integer number.
It is important that one can always reduce the system to normalized Fuchsian form in a given regular singular point using the transformation \eqref{eq:transformation}, Refs.  \cite{horn1891theorie,moser1959order}.
Moreover the systems which appear in multiloop calculations usually can be reduced to global normalized Fuchsian form \cite{Lee:2017oca}, i.e. the normalized Fuchsian form at all singular points. In this case the matrix $M(x)$ has the form \eqref{eq:fuchsian} with all matrix residues $M_a$ ($a\in S$) being non-resonant.

For simplicity of presentation we will assume below that the system \eqref{eq:de} is already in global normalized Fuchsian form.

Then it has a generalized series solution in the neighborhood of singular point $a$:
\begin{equation}
	F(x)=U(x,\underline{a}) = H^{a}(x)(x-a)^{M_a},\qquad
	H^{a}(x)= \sum_{n=0}^{\infty} H^{a}_n\cdot (x-a)^n,\label{eq:frobenius}
\end{equation}
where $H^{a}_n$ are constant matrices and $H^{a}_0=I$ is the identity matrix, see. e.g., \cite{haraoka2020linear}. The series converges in a disk $D_a$ with the radius being the distance to the closest singular point. Note that in order to understand this expansion as a single-valued function, one has to fix the position of the cut and the branch of the function $(x-a)^{M_a} = \exp[M_a \ln (x-a)]$. In all our examples we choose the cut as a ray going from the singular point downward.

Let us now describe the monodromy representation in terms of these  generalized series solutions.
First, from the form \eqref{eq:frobenius} we see that the generators $\rho(\gamma_a)=\M_{a}$, up to similarity, are determined by the matrix residues. We have
\begin{equation}
	\M_{a} = C_{a}^{-1}e^{2\pi i M_{a}}C_{a},\label{eq:local}
\end{equation}
where the matrices $C_{a}$ depend, in particular, on the choice of the base point $b$ and the fundamental matrix $F(x)$. Therefore, in order to define the monodromy representation it remains ``only'' to find the matrices $C_{a_i}$.

Let us fix a singular point $a_0$ and choose the fundamental matrix $F(x)$ defined via
\begin{equation}
	F(x) = U(x,\underline{a_0})
\end{equation}
in the convergence disk $D_{a_0}$ of series in Eq. \eqref{eq:frobenius}. Let us also choose $b=b_0\in D_{a_0}$. For this choice it is obvious that $C_{a_0}=I$ in Eq. \eqref{eq:local} and, therefore, $\M_{a_0} = e^{2\pi i M_{a_0}}$.

Suppose the convergence disk  $D_{a_1}$ of the generalized series expansion at $x=a_1$ has a non-empty intersection with $D_{a_0}$. Then in this intersection we can write
\begin{equation}
	F(x) = U(x,\underline{a_1})U^{-1}(x,\underline{a_1})U(x,\underline{a_0})
	\label{eq:F_c}
\end{equation}

\begin{figure}
	\centering
	\includegraphics[width=0.7\linewidth]{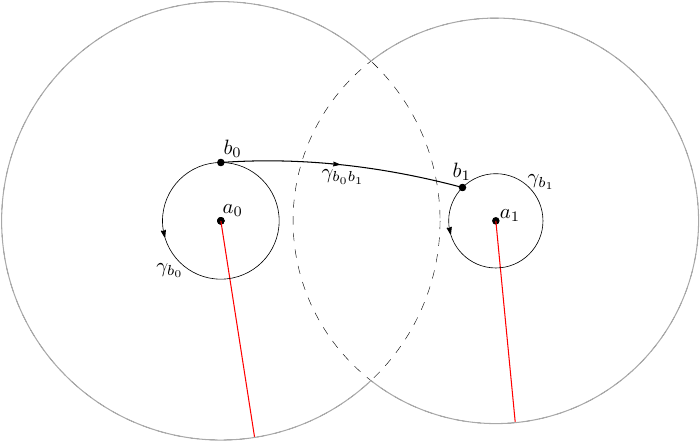}
	\caption{Overlapping disks $D_{a_0}$ and $D_{a_1}$ and the generators $\gamma_{a_0}=\gamma_{b_0}$ and $\gamma_{a_1}=\gamma_{b_0b_1}\gamma_{b_1}\gamma_{b_1b_0}$, where $\gamma_{b_1b_0}$ denotes the path opposite to $\gamma_{b_0b_1}$. The red lines (cuts) denote the gluing with previous and next sheets.}
	\label{fig:contours}
\end{figure}

Since the two fundamental matrices $U(x,\underline{a_0})$ and $U(x,\underline{a_1})$ may differ only by a constant matrix multiplication from the right, we conclude that $U^{-1}(x,\underline{a_1})U(x,\underline{a_0})$ does not depend on $x$. Then, Eq. \eqref{eq:F_c} allows one to analytically continue $F(x)$ onto $D_{a_1}$. The matrix
\begin{equation}
	U(\underline{a_1},\underline{a_0})\stackrel{\text{def}}{=}U^{-1}(x,\underline{a_1})U(x,\underline{a_0})\,,
\end{equation}
which relates the fundamental solutions in the vicinities of points $a_0$ and $a_1$, is a particular case of what is called the connection matrix \cite{haraoka2020linear}.
Then it is easy to see that we can take
\begin{equation}
	C_{a_1}=U(\underline{a_1},\underline{a_0})\,.
\end{equation}
Indeed, we have
\begin{equation}
	F(x)=U(x,\underline{a_0}) \stackrel{\gamma_{b_0b_1}}{\longrightarrow}
	U(x,\underline{a_1}) C_{a_1}
	\stackrel{\gamma_{b_1}}{\longrightarrow}
	U(x,\underline{a_1}) e^{2\pi i M_c} C_{a_1}
	\stackrel{\gamma_{b_1b_0}}{\longrightarrow}
	U(x,\underline{a_0}) C_{a_1}^{-1} e^{2\pi i M_{a_1}} C_{a_1},
\end{equation}
where by $\stackrel{\gamma}{\longrightarrow}$ we denote the analytical continuation of $F(x)$ along the path $\gamma$ and the paths are depicted in Fig. \ref{fig:contours}.

Applying repeatedly the same approach to the analytical continuation into the vicinity of $a_i$, we conclude that
\begin{equation}
	C_{a_i}=U(\underline{a_i},\underline{a_{i-1}})\ldots U(\underline{a_2},\underline{a_{1}})
	U(\underline{a_1},\underline{a_{0}})\,.
	\label{eq:Tres}
\end{equation}

Our method is based on the high-precision (at least with hundreds of decimal digits) evaluation of the connection matrices $U(\underline{a_i},\underline{a_{i-1}})$ which are then used for the determination of $C_{a_i}$ and $\M_{a_i}$ via Eqs. \eqref{eq:Tres} and \eqref{eq:local}. For this purpose we can use the approach described in Ref. \cite{Lee:2017qql} which allows one to derive the fixed-order recurrence relations between the coefficients of the generalized power series \eqref{eq:frobenius} thus making the computational complexity of evaluating $N$ terms to be $O(N)$. In particular, this approach is implemented in \texttt{Libra} \cite{Lee:2020zfb} via functions \texttt{SeriesSolutionData} and \texttt{ConstructSeriesSolution}.

\section{Monodromy in $d$ dimensions}\label{sec:d}

Using this high-precision results and the integer relation algorithms one may try to recognize the analytical form of the monodromy matrices. However in order to perform this recognition, one has to rely on some ansatz for the functions (of $d$) which may appear.
One has to take into account the freedom in the choice of  a basis of solutions which corresponds to transformations $F(x)\to F(x)C$ with constant matrix $C=C(d)$. Under this transformation the monodromy generators transform as $\M_a\to C^{-1}\M_a C$.

In this section we discuss the properties of the monodromy representation which are specific for the differential systems which appear in multiloop calculations in dimensional regularization.

\subsection{Dimension shifts}
Let us show that for such systems there exists a basis for which the monodromy matrices are periodic functions of $d$.
The multiloop integrals satisfy the differential system
\begin{equation}
	\partial_x J(x,d) = M(x,d) J(x,d), \label{eq:de1}
\end{equation}
and the dimensional recurrence relation
\begin{equation}
	J(x,d+2)=L(x,d)J(x,d),\label{eq:drr}
\end{equation}
where the matrices $M(x,d)$ and  $L(x,d)$ are rational in $x$ and $d$.
We assume that Eqs. \eqref{eq:de1} and \eqref{eq:drr} are compatible, i.e. there is a fundamental solution $F(x)$ of \eqref{eq:de1} which also satisfies \eqref{eq:drr}:
\begin{equation}
	F(x,d+2) = L(x,d)F(x,d).\label{eq: Ftilde1}
\end{equation}
Upon the analytical continuation along a loop $\gamma$, this equation reads
\begin{equation}
	F(x,d+2)\rho(\gamma,d+2) = L(x,d)F(x,d)\rho(\gamma,d).\label{eq: Ftilde2}
\end{equation}
Here we have taken into account that $L(x,d)$ does not change upon the analytical continuation being a rational function of $x$. Then, from Eqs. \eqref{eq: Ftilde1} and \eqref{eq: Ftilde2} we have the periodicity condition
\begin{equation}
	\rho(\gamma,d) = \rho(\gamma,d+2).
\end{equation}

Therefore, indeed, there exists a basis for which the monodromy matrices are periodic functions of $d$. In other words, there is a basis for which the monodromy matrices are functions of $z=\exp(i\pi d)$. Moreover, it is easy to see that the eigenvalues of the monodromy generators are proportional to rational powers of $z$. We prove this statement in Appendix \ref{sec:appendix}.

It appears that for all examples considered in this paper it \add{is} possible to find a basis (or the matrix $C(d)$) for which the monodromy group $\M$ is a subgroup of $GL(n,\mathbb{Z}[z,1/z])$, i.e. a group of matrices with elements being the Laurent polynomials of $z$ with rational coefficients. Let us note that the monodromy representation which maps the fundamental group into $GL(n,\mathbb{Z}[z,1/z])$ appeared earlier in Ref. \cite{shimada1996picard}.

\subsection{Twisted Riemann bilinear relations}

Another specific feature of the differential systems for Feynman integrals is the existence of bilinear relations between the fundamental matrices $F(x,d)$ and $F(x,-d)$ corresponding to an irreducible block of the differential system \cite{cho1995intersection,aomototheory,Broadhurst:2018tey,Lee:2018jsw,lee2019differential,Lee2020,duhr2025twisted}. These blocks usually correspond to maximal cuts of specific sectors in the integral family. The relations (called \emph{twisted Riemann bilinear relations}) have the form
\begin{equation}
	F^{\intercal}(x,-d) D(x,d) F(x,d) = B(d),\label{eq:trbr}
\end{equation}
where $D(x,d)$ and $B(d)$ are the matrices related to the intersection forms of twisted cocycles and cycles, respectively \cite{cho1995intersection,aomototheory,Broadhurst:2018tey,duhr2025twisted}. It is important that $D(x,d)$ is a non-degenerate rational matrix and the matrix $B(d)$ does not depend on $x$.

Let us analytically continue this relation along some closed contour $\gamma$. We have
\begin{equation}
	\rho^{\intercal}(\gamma,-d) F^{\intercal}(x,-d) D(x,d) F(x,d) \rho(\gamma,d)= B(d),
\end{equation}
which leads to the relation
\begin{equation}
	\rho^{\intercal}(\gamma,-d) B(d) \rho(\gamma,d)= B(d).\label{eq:qmondromy}
\end{equation}
It means that the matrices $\rho(\gamma,d)$ and $\rho^{\intercal-1}(\gamma,-d)$ are similar. We stress that the similarity matrix $B(d)$ does not depend on the contour $\gamma$.

Two remarks are in order. The first remark is that the relations \eqref{eq:qmondromy} are not only the consequence of twisted Riemann bilinear relations \eqref{eq:trbr} but also imply their existence, i.e., imply the existence of the rational matrix $D(x,d)$. Indeed, let us consider the quantity
\begin{equation}
	\widetilde{D}(x,d)=F^{\intercal-1}(x,-d) B(d) F^{-1}(x,d)\label{eq:D}
\end{equation}
After analytical continuation along the loop $\gamma^{-1}$, it turn into
\begin{equation}
	F^{\intercal-1}(x,-d) \rho^{\intercal}(\gamma,-d) B(d) \rho(\gamma,d) F^{-1}(x,d),
\end{equation}
which, by Eq. \eqref{eq:qmondromy} is equal to $\widetilde{D}(x,d)$. It means that $\widetilde{D}(x,d)$ has trivial monodromy, thus being a rational function of $x$.

Second remark is that for irreducible monodromy representation Eq. \eqref{eq:qmondromy} has a unique, up to an overall factor, nonzero solution which is necessarily invertible, which follows from Schur's lemma. Since  $B^{\intercal}(-d)$ satisfies the same equation \eqref{eq:qmondromy} as $B(d)$, those two matrices coincide also up to an overall factor.
We can fix the overall factor in the definition of $B(d)$ by requiring that the elements of $B$ are polynomial in $z$ with common greatest divisor equal to unity. Then it is easy to see that with this requirement $B^{\intercal}(-d)=\pm z^{-k} B(d)$, where $k\in\mathbb{Z}_+$.

Besides we find that in all cases the  determinant of the matrix $B(d)$ is a constant multiple of a product of cyclotomic polynomials of $z$ and of some power of $z$. It is easy to see that for any zero of these cyclotomic polynomials\footnote{Recall that cyclotomic polynomials are irreducible factors of $z^n-1$ and therefore all their zeros are roots of unity.} (when $d$ is rational) the found monodromy representation becomes reducible.
Indeed, for such values of $z$ the kernel of $B(d)$ is a proper subspace of $\mathbb{C}^n$ (i.e. $0\neq \ker B(d)\neq \mathbb{C}^n$) thanks to our choice of the common factor in $B(d)$. On the other hand, rewriting Eq. \eqref{eq:qmondromy} as
\begin{equation}
	B(d) \rho(\gamma,d)= \rho^{\intercal-1}(\gamma,-d)B(d),
\end{equation}
we can easily see that $\ker B(d)$ is the invariant subspace of $\rho(\gamma,d)$ for all loops $\gamma$. Indeed, mutiplying this equation by $u\in\ker B(d)$, we find that $B(d) \rho(\gamma,d)u=0$, i.e. $\rho(\gamma,d)u\in \ker B(d)$.

\section{Finding monodromy as function of $d$}\label{sec:algorithm}
Our heuristic approach consists of the following steps:
\begin{enumerate}
	\item We pick a numeric $d$ such that $z=\exp(i\pi d)$ is transcendental. E.g., thanks to the Lindemann theorem one can take $d=r/\pi$, where $r\neq 0$ is a rational number.
	\item Then we find sufficiently many terms of generalized power series expansions at each singular point. If necessary, we add also the expansions around some regular points, so that the union of the convergence disks is a connected set.
	\item Using these expansions, we construct numeric matrices of monodromy generators along the lines described in the previous section.
	\item We pick a vector $v_1$ corresponding to a non-degenerate eigenvalue $\propto z^k$ ($k\in \mathbb{Z}$) of one of the monodromy generators. Acting on this vector by various elements of monodromy group we obtain a basis.
	\item We transform the monodromy generators to this basis by applying the similarity transformation $\M_a\to C^{-1}\M_a C$ with $C$ consisting of these vectors as columns and try to recognize their matrix elements as rational functions of $z$ using the integer relation algorithms (such as PSLQ \cite{ferguson1999analysis}).
	\item If the recognized generators are in $GL(n,\mathbb{Q}(z))$ but not in $GL(n,\mathbb{Z}[z,1/z])$, we succeed in finding additional similarity transformation which maps the generators into $GL(n,\mathbb{Z}[z,1/z])$.
\end{enumerate}

Note that if one assumes the existence of a basis where $\M\subset GL(n,\mathbb{Q}(z))$, step 4 will necessarily provide a possible basis for such a representation. Indeed, it is easy to see that the transformation matrix between those two bases will also belong to $GL(n,\mathbb{Q}(z))$ (up to an overall arbitrary factor depending on the normalization of $v_1$).

For the recognition of the rational functions of $z$  at step 5 we use the following standard approach. Suppose that $f$ is a rational function with rational coefficients to be recognized.  Let $f(z)$ is its high-precision numerical value at transcendental argument $z$ from step 1 above. Then we search for the nontrivial integer relation between the numbers
\begin{equation}
	1,z,\ldots, z^k, f(z), zf(z),\ldots, z^k f(z),
\end{equation}
where $k$ is chosen to be sufficiently large. If such a relation exists it allows one to express $f(z)$ as a ratio of two polynomials of $z$ with integer coefficients.

\section{Examples}\label{sec:examples}
All examples presented in this section as well as several additional examples are included in the ancillary files as \textit{Mathematica} notebooks.
\subsection{Example 1: two-loop sunrise}
Let us explain our method on the example of equal-mass sunrise master integrals $J=(j_0,j_1,j_2)^\intercal$, where
\begin{equation}
	j_k(x) = \int \frac{dl_1\,dl_2/(i\pi^{d/2})^2}{(1-l_1^2-i0)(1-l_2^2-i0) \left[1-(q-l_1-l_2)^2-i0\right]^k}\,,\quad q^2=x^{-1}\,.
\end{equation}
These functions are solutions to the system \eqref{eq:de} with
\begin{equation}
	M(x)=
	\left(
	\begin{array}{ccc}
		0 & 0 & 0 \\
		0 & -\frac{d-3}{x} & -\frac{3}{x} \\
		\frac{(d-2)^2}{2 (x-1) (9 x-1)} & \frac{(d-3) (3 d-8) (3 x-1)}{2 (x-1) (9 x-1)} & \frac{9 (3d-8) x^2+10 (2-d)x+4-d}{2 (x-1) x (9 x-1)} \\
	\end{array}
	\right)
\end{equation}
Note that this system is in global normalized Fuchsian form with $S=\{0,1,1/9,\infty\}$. Therefore, the monodromy generator $\M_{a}$ corresponding to a loop around $a$ is similar to $\exp(2\pi i M_{a})$. We have
\begin{gather}
	\M_0 \sim e^{2\pi i M_0}\sim \diag(1, z^{-1}, z^{-2}),\quad
	\M_{1/9} \sim e^{2\pi i M_{1/9}}\sim\diag(1, 1, z^2),\nonumber\\
	\M_1 \sim e^{2\pi i M_{1}} \sim \diag(1, 1, z^2),\quad
	\M_\infty \sim e^{2\pi i M_{\infty}} \sim\diag(1, 1, z^{-1}).
\end{gather}

We take $d=12/\pi$ and calculate with high precision the connection matrices
\begin{align}
	U(\underline{1/9},\underline{0})&=
	U^{-1}(x_{1},\underline{1/9})U(x_{1},\underline{0}),\label{eq:x1}\\
	U(\underline{1},\underline{1/9})&=
	U^{-1}(x_{2},\underline{1})U(x_{2},\underline{1/9}),\label{eq:x2}
\end{align}
where $x_{1}$ and $x_{2}$ should be chosen in the intersection of the corresponding convergence disks. We do this by constructing the generalized series representation for the quantities in the right-hand sides of Eqs. \eqref{eq:x1} and \eqref{eq:x2} using \texttt{Libra}. We choose $x_1=1/18$ and $x_2=17/81$ in order to optimize the convergence rate of the two series which depend on the corresponding point.

Using these connection matrices, we obtain numerically the monodromy generators as
\begin{gather}
	\M'_0=e^{2\pi i M_0},\quad
	\M'_{1/9}=U^{-1}(\underline{1/9},\underline{0})e^{2\pi i M_{1/9}}U(\underline{1/9},\underline{0}),\\	\M'_{1}=U^{-1}(\underline{1},\underline{0})
	e^{2\pi i M_{1}}U(\underline{1},\underline{0}),\quad
	\M'_{\infty}=\left(\M'_1\M'_{1/9}\M'_0\right)^{-1},
\end{gather}
where $U(\underline{1},\underline{0})=U(\underline{1},\underline{1/9})U(\underline{1/9},\underline{0})$. Note that by construction the generators $\M'_a$ are block-triangular, similar to the matrix $M(x)$.
Next we choose the basis as follows. We take $v_1$ to be the eigenvector of $\M_0'$ with the eigenvalue $1$. The two other vectors of the basis are obtained as\footnote{The subtraction of the vector $v_1$ in both cases preserves the block-triangular form of the monodromy matrices.}
\begin{equation}
	v_2=\M_{1/9}'v_1-v_1,\quad v_3=\M_{1}'v_1-v_1\,.
\end{equation}
Applying the similarity transformation with $C=(v_1|v_2|v_3)$ and recognizing rational functions of $z$, we obtain $\M''_a=C^{-1}\M'_a C$:
\begin{gather}
	\M''_0=\left(
	\begin{array}{ccc}
		1 & 0 & 0 \\
		0 & \frac{3 z+1}{z^2} & \frac{(2 z+1)^2}{z^2} \\
		0 & -\frac{3}{z (2 z+1)} & -\frac{2}{z} \\
	\end{array}
	\right),\quad
	\M''_{1/9}=\left(
	\begin{array}{ccc}
		1 & 0 & 0 \\
		1 & z^2 & z (z+1) (2 z+1) \\
		0 & 0 & 1 \\
	\end{array}
	\right),\nonumber\\
	\M''_{1}=\left(
	\begin{array}{ccc}
		1 & 0 & 0 \\
		0 & 1 & 0 \\
		1 & -\frac{3 (z+1)}{2 z+1} & z^2 \\
	\end{array}
	\right),\quad
	\M''_{\infty}=
	\left(
	\begin{array}{ccc}
		1 & 0 & 0 \\
		-\frac{1}{z} & \frac{z+3}{z} & \frac{2 z+1}{z} \\
		\frac{z+2}{z (2 z+1)} & -\frac{3 (z+2)}{z (2 z+1)} & -\frac{2}{z} \\
	\end{array}
	\right)\,.
\end{gather}
Applying the simple diagonal transformation $\widetilde{C}=\diag(1,1,(1+2z)^{-1})$, we obtain the monodromy generators $\M_a=\widetilde{C}^{-1}\M''_a\widetilde{C}$ as elements of $GL(n,\mathbb{Z}[z,1/z])$:
\begin{gather}
	\M_0(z)=\left(
	\begin{array}{ccc}
		1 & 0 & 0 \\
		0 & \frac{3 z+1}{z^2} & \frac{2 z+1}{z^2} \\
		0 & -\frac{3}{z} & -\frac{2}{z} \\
	\end{array}
	\right),\quad
	\M_{1/9}(z)=\left(
	\begin{array}{ccc}
		1 & 0 & 0 \\
		1 & z^2 & z (z+1)\\
		0 & 0 & 1 \\
	\end{array}
	\right),\nonumber\\
	\M_{1}(z)=\left(
	\begin{array}{ccc}
		1 & 0 & 0 \\
		0 & 1 & 0 \\
		2 z+1 & -3 (z+1) & z^2 \\
	\end{array}
	\right),\quad
	\M_{\infty}(z)=
	\left(
	\begin{array}{ccc}
		1 & 0 & 0 \\
		-\frac{1}{z} & \frac{z+3}{z} & \frac{1}{z} \\
		\frac{z+2}{z} & -\frac{3 (z+2)}{z} & -\frac{2}{z} \\
	\end{array}
	\right)\,.
\end{gather}

Let us now consider the bilinear relation \eqref{eq:qmondromy}. The monodromy matrices have the block-triangular form $\M_a=\begin{pmatrix} 1 &0\\ *& \widetilde{\M}_a
\end{pmatrix}$. Then for the blocks $\widetilde{\M}_a$ we have
\begin{equation}
	\widetilde{\M}_a^{\intercal}(1/z)B(d)\widetilde{\M}_a(z)=B(d), \qquad a=0,1/9,1,\infty.
\end{equation}
Solving these equations as a linear system for the elements of the matrix $B$, we find
\begin{equation}
	B(d)=
	\begin{pmatrix}
		3 (z-1) & 3 z \\
		-3 & z-1 \\
	\end{pmatrix},\qquad \det B(d) =3(z^2+z+1).
\end{equation}
The matrix $B(d)$ is degenerate at $z=\exp(\pm 2\pi i/3)$.
\subsection{Example 2: three-loop forward box}
\begin{figure}
	\centering
	\includegraphics[width=0.5\linewidth]{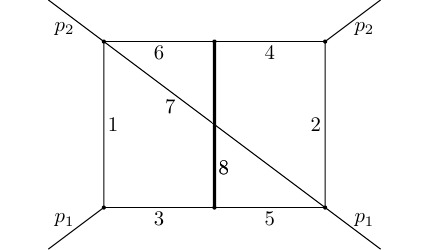}
	\caption{Three-loop forward box from Ref. \cite{Mistlberger:2018etf}. Here $p_1^2=p_2^2=0$, $(p_1+p_2)^2=x$, thin lines correspond to massless propagators and thick line corresponds to massive propagator with $m^2=1$.}
	\label{fig:mstl}
\end{figure}

Our next example is concerned with the elliptic sector from Ref. \cite{Mistlberger:2018etf}. We consider the integrals of the family
\begin{equation}
	j(n_1,\ldots,n_8|x) = \int \frac{dl_1\,dl_2\,dl_3}{(\pi^{d/2})^3}\prod_{k=1}^8{\Im (D_k+i0)^{-n_k}}.
\end{equation}
which corresponds to the maximal cut of the diagram in Fig \ref{fig:mstl}.

We can choose four master integrals of this family as
\begin{gather}
	j_1(x)=j(11111111|x),\quad
	j_2(x)=j(11111112|x),\nonumber\\
	j_3(x)=j(11112111|x),\quad
	j_4(x)=j(21111111|x).
\end{gather}
The column $j=(j_1,\ldots,j_4)^\intercal$ satisfy the differential system $\partial_x j = \widetilde{M}(x) j$ with
\begin{equation}
	\widetilde{M}(x)=\frac{\widetilde{M}_{0}}{x}+
	\frac{\widetilde{M}_{a_+}}{x-a_+}+
	\frac{\widetilde{M}_{a_-}}{x-a_-}+
	\frac{\widetilde{M}_{-8}}{x+8},
\end{equation}
where $a_{\pm}=\frac12(-11\pm5\sqrt{5})$ are the zeros of the polynomial $x^2+11x-1$. There are two issues about this system. First, the singularity at $x=-8$ is apparent. Second, the matrix $\widetilde{M}_{0}$ has resonant eigenvalues $d/2-4$ and $d/2-3$. We can fix both issues by a rational transformation $j=TJ$ (recall that the monodromy group is not changed by this transformation). We obtain the system in the normalized global Fuchsian form
\begin{equation}
	\partial_x J = M(x),\qquad M(x)=\frac{M_0}{x}+
	\frac{M_{a_+}}{x-a_+}+
	\frac{M_{a_-}}{x-a_-}.
\end{equation}
Then we have
\begin{gather}
	\M_0\sim \left(
	\small\begin{array}{cccc}
		z & 1 & 0 & 0 \\
		0 & z & 0 & 0 \\
		0 & 0 & 1/z & 0 \\
		0 & 0 & 0 & 1/z^{2} \\
	\end{array}
	\right),\quad
	\M_\infty \sim \left(
	\small
	\begin{array}{cccc}
		{1}/{z^2} & 1 & 0 & 0 \\
		0 & {1}/{z^2} & 0 & 0 \\
		0 & 0 & 1 & 0 \\
		0 & 0 & 0 & {1}/{z^3} \\
	\end{array}
	\right),\nonumber\\
	\M_{a_\pm} \sim \diag(1, 1, 1, z^4).
\end{gather}
Again, we take $d=12/\pi$. Note that convergence radius of the series at $x=0$ is determined by the distance to the point $x=a_+\approx 0.09$. Meanwhile, the distance to the point $x=a_-\approx -11.09$ is two orders of magnitude larger. Therefore, in order to increase the convergence rate we consider also the series around a few intermediate points in the interval $(a_-,0)$. We take 5 additional points $-37/9, -29/19, -5/9, -1/5, -1/13$. Then the connection matrix $U(\underline{a_-},\underline{0})$ is constructed as
\begin{align}
	U(\underline{a_-},\underline{0})=
	U\left(\underline{a_-},-\underline{\tfrac{37}{9}}\right)
	U\left(-\underline{\tfrac{37}{9}},-\underline{\tfrac{29}{19}}\right)
	U\left(-\underline{\tfrac{29}{19}},-\underline{\tfrac{5}{9}}\right)
	U\left(-\underline{\tfrac{5}{9}},-\underline{\tfrac{1}{5}}\right)
	U\left(-\underline{\tfrac{1}{5}},-\underline{\tfrac{1}{13}}\right)
	U\left(-\underline{\tfrac{1}{13}},\underline{0}\right).
\end{align}

Similar to the previous example we obtain numerically the monodromy generators as
\begin{gather}
	\M'_0=e^{2\pi i M_0},\quad
	\M'_{a_\pm}=U^{-1}(\underline{a_\pm},\underline{0})e^{2\pi i M_{a_\pm}}U(\underline{a_\pm},\underline{0}),\quad
	\M'_{\infty}=\left(\M'_{a_+}\M'_{0}\M'_{a_-}\right)^{-1}.
\end{gather}

Next we choose a basis as follows. We take $v_1$ to be the eigenvector of $\M_0'$ with the eigenvalue $z$. The three other vectors of the basis are chosen as\footnote{In fact, the denominator $1+z^2$ in the definition of $v_3$ was obtained as a result of additional transformation similar to $\widetilde{C}$ in the previous example.}
\begin{equation}
	v_2=\M_{a_+}'v_1-v_1,\quad v_3=\frac{\M_{a_-}'v_1-v_1}{1+z^2},\quad
	v_4=\M_{0}'(v_2-v_3)\,.
\end{equation}
Applying the similarity transformation with $C=(v_1|v_2|v_3|v_4)$ and recognizing rational functions of $z$, we obtain $\M_a=C^{-1}\M'_aC$:
\begin{gather}
	\M_0=\left(
	\begin{array}{cccc}
		z & (z-1) z & (z-1) z & 1-z \\
		0 & \frac{(z+1) \left(z^2+1\right)}{z^2} & \frac{(z+1) \left(z^2+1\right)}{z^2} & -\frac{z^3+z^2+1}{z^3} \\
		0 & -1 & -1 & \frac{1}{z} \\
		0 & 1 & 0 & 0 \\
	\end{array}
	\right),\\
	\M_{a_+}=\left(
	\begin{array}{cccc}
		1 & 0 & 0 & 0 \\
		1 & z^4 & z^2 \left(z^3+z^2+z+2\right) & 0 \\
		0 & 0 & 1 & 0 \\
		0 & 0 & 0 & 1 \\
	\end{array}
	\right),\\
	\M_{a_-}=\left(
	\begin{array}{cccc}
		1 & 0 & 0 & 0 \\
		0 & 1 & 0 & 0 \\
		z^2+1 & -\frac{2 z^3+z^2+z+1}{z} & z^4 & \frac{z^5+z^3+2 z^2+1}{z} \\
		0 & 0 & 0 & 1 \\
	\end{array}
	\right),\\
	\M_{\infty}=\left(
	\begin{array}{cccc}
		\frac{1}{z} & 0 & z-1 & 0 \\
		0 & 0 & 0 & 1 \\
		\frac{z^3-z^2+z+1}{z^5} & -\frac{(z+1) \left(z^2-z+2\right)}{z^5} & \frac{z^3-z^2+2 z+1}{z^3} & \frac{(z+1) \left(2 z^2-z+1\right)}{z^5} \\
		-\frac{1}{z^2} & \frac{1}{z^2} & -1 & 0 \\
	\end{array}
	\right).
\end{gather}

Similar to the previous example, let us consider the relations
\begin{equation}
	\M_a^{\intercal}(1/z)B(d)\M_a(z)=B(d), \qquad a=0,a_+,a_-,\infty.
\end{equation}
Solving these equations we find the matrix $B$ as rational function of $z$:
\begin{gather}
B(d)=\left(
\begin{array}{cccc}
	0 & z^5 & z^3 \left(z^2+1\right) & -z^4 \\
	-z & z \left(1-z^4\right) & -z^3 \left(z^3+z^2+z+2\right) & 0 \\
	-z \left(z^2+1\right) & 2 z^3+z^2+z+1 & z \left(1-z^4\right) & -z^5-z^3-2 z^2-1 \\
	z^2 & 0 & z \left(z^5+2 z^3+z^2+1\right) & \left(z^4-1\right)\left(z^2-z+1\right)  \\
\end{array}
\right),\nonumber\\
\det B(d)=z^4(1-z)  \left(1+z^2\right) \left(1-z^4\right)^2 \left(1-z^5\right)\,.
\end{gather}

\subsection{Example 3: three-loop off-shell massless vertex}

\begin{figure}
	\centering
	\includegraphics[width=0.5\linewidth]{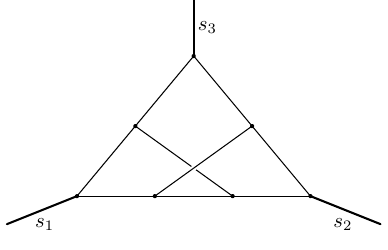}
	\caption{Three-loop off-shell massless vertex.}
	\label{fig:vertex3}
\end{figure}
While our previous consideration concerned the case of one variable, the same approach can be applied to the systems with respect to several variables. Note that for Feynman integrals there is a natural compactification of the space of kinematic variables which can be easily found by the dimensional analysis. The integrals depend nontrivially only on the dimensionless ratios of the kinematic invariants. In particular, when all kinematic invariants have the same dimension (equal to $[m^2]$), the integrals depend nontrivially on the point in the complex projective space $\mathbb{CP}^{k}$. E.g., our previous consideration concerned the case of $\mathbb{CP}^{1}=\overline{\mathbb{C}}$.

\begin{figure}
	\centering
	\includegraphics[width=0.6\linewidth]{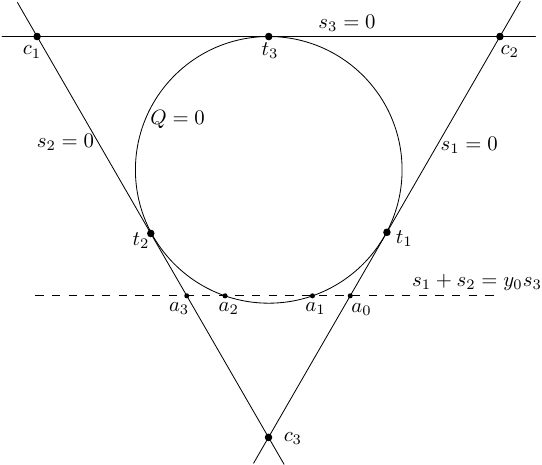}
	\caption{Picture of the reducible variety (algebraic curve) defined by the equation $s_1s_2s_3 Q=0$ on the real section of the chart $s_1+s_2+s_3=1$.
	This curve has six singular points. Three of them, $c_{1,2,3}$, are intersection points (crunodes in the terminology of Ref. \cite{ponzano1969monodromy})  and other three, $t_{1,2,3}$, are the points of tangency (tacnodes). The points $t_{1}$ and $t_{2}$ belong to the line $s_1+s_2=s_3$.
	The dashed line depicts the ``generic'' section which we use to obtain the monodromy generators. The four finite singular points are denoted as $a_{0,1,2,3}$, while the fifth point corresponding to the intersection with the line $s_3=0$, is $a_4=\infty$.}
	\label{fig:singularities}
\end{figure}

Now we examine the maximal cut of the three-loop off-shell massless vertex shown in Fig. \ref{fig:vertex3}. This family depends on three variables $s_1, s_2, s_3$ which can be considered as homogeneous coordinates in $\mathbb{CP}^{2}$.
We consider the chart $s_3=1$.
Then the four master integrals $\boldsymbol{j}=(j_1,\ldots, j_4)^\intercal$ in the top sector satisfy
\begin{equation}
	\partial_{s_1} \boldsymbol{j} =M_{1}(s_1,s_2)\boldsymbol{j},\quad
	\partial_{s_2} \boldsymbol{j} =M_{2}(s_1,s_2)\boldsymbol{j}\,.
\end{equation}
Understanding this system as defined on $\mathbb{CP}^{2}$, we find that the singular locus is the algebraic curve $L$ determined by the equation $s_1s_2s_3 Q=0$, where
\begin{equation}
	Q = s_1^2+s_2^2+s_3^2-2s_1s_2-2s_1s_3-2s_2s_3\,.
\end{equation}
It is convenient to depict these singularities on the real section of the chart $s_1+s_2+s_3=1$, see Fig. \ref{fig:singularities}.

In order to find the monodromy generators we pass to the variables $x=s_1-s_2$ and $y=s_1+s_2$. Then we consider a generic section $y=y_0$, where $1/2<y_0<1$. Here ``generic'' means that this section has maximal possible number of intersections with  $L$, which coincides with the degree of the defining polynomial $s_1s_2s_3 Q$. According to Picard-Severi theorem \cite{pham1967singularites,ponzano1969monodromy}, the generators of the fundamental group can be chosen to lie in such a section. In the differential system with respect to $x$ we have the following singular points:
\begin{align}
	a_0=-y_0,\quad a_1=-\sqrt{2y_0-1},\quad a_2=\sqrt{2y_0-1}, \quad a_3=y_0,
	\quad a_4=\infty\,.
\end{align}

Using our method, we obtain the following monodromy generators:
\begin{gather}
	\M_{s_1}=\M_{-y_0}=\left(
	\begin{array}{cccc}
		1 & 0 & z & \frac{1}{z} \\
		z^4-z^3 & z^3 & z & z^3 \\
		0 & 0 & z^3+1 & z \\
		0 & 0 & -z^2 & 0 \\
	\end{array}
	\right),\quad
	\M_{Q}=\M_{\pm\sqrt{2y_0-1}}=\left(
	\begin{array}{cccc}
		z^2 & 0 & 0 & z+1 \\
		0 & -\frac{1}{z} & 0 & 0 \\
		0 & 0 & -\frac{1}{z} & 0 \\
		0 & 0 & 0 & -\frac{1}{z} \\
	\end{array}
	\right),
	\nonumber\\
	\M_{s_2}=\M_{y_0}=\left(
	\begin{array}{cccc}
		1 & -z^2 & 0 & -z^4 \\
		0 & 0 & 0 & -z^2 \\
		\frac{z-1}{z} & -z^2 & z^3 & -1 \\
		0 & z & 0 & z^3+1 \\
	\end{array}
	\right),\quad
	\M_{s_3}=\M_{\infty}=\left(
	\begin{array}{cccc}
		\frac{1}{z^3} & \frac{1}{z^2} & -\frac{1}{z^3} & 0 \\
		0 & \frac{z+1}{z^3} & -\frac{1}{z^3} & 0 \\
		0 & \frac{1}{z^2} & 0 & 0 \\
		\frac{1-z}{z^3} & -\frac{1}{z^3} & \frac{1}{z^2} & \frac{1}{z^2} \\
	\end{array}
	\right).
\end{gather}

We have four distinct generators, each corresponding to one irreducible component of the curve $L$. Being the images of the corresponding generators of the fundamental group, in addition to the relation  $\M_{s_3}\M_{s_2}\M_{Q}^2\M_{s_1}=1$ they should at least satisfy the six relations\footnote{Some relations in the set \eqref{eq:relations} are redundant, in particular, relations $c_1,c_2,t_3$ can be derived from $t_1,t_2,s_3$.} which appear for each singular point $c_i$ and $t_i$. Indeed, we have
\begin{gather}
	c_1:\quad [\M_{s_3},\M_{Q}^{-1}\M_{s_2}\M_{Q}]=0,\nonumber\\
	c_2:\quad [\M_{s_3},\M_{Q}\M_{s_1}\M_{Q}^{-1}]=0,\nonumber\\
	c_3:\quad [\M_{s_1},\M_{Q}^{-1}\M_{s_2}\M_{Q}]=0,\nonumber\\
	t_1:\quad [\M_{s_1},\M_{Q}\M_{s_1}\M_{Q}]=0,\nonumber\\
	t_2:\quad [\M_{s_2},\M_{Q}\M_{s_2}\M_{Q}]=0,\nonumber\\
	t_3:\quad [\M_{s_3},\M_{Q}\M_{s_3}\M_{Q}]=0.\label{eq:relations}
\end{gather}

Again, solving the equations
\begin{equation}
	\M_c^{\intercal}(1/z)B(d)\M_c(z)=B(d), \qquad c=s_1,s_2,s_3, Q.
\end{equation}
we find
\begin{equation}
	B(d) = \left(
	\begin{array}{cccc}
		-z^2+z-1 & 0 & 0 & -z \\
		0 & 0 & z^2 & 1 \\
		0 & 1 & 0 & z^2 \\
		-z & z^2 & 1 & z \\
	\end{array}
	\right),\qquad \det B(d) =-(z-1) \left(z^2+1\right) \left(z^5-1\right)\,.
\end{equation}

\section{Conclusion}\label{sec:conclusion}

In the present paper we have introduced a practical method to find the monodromy group of the differential systems for multiloop integrals in $d$ dimensions. On several examples we have shown that this group is isomorphic to a subgroup of $GL(n,\mathbb{Z}[z,1/z])$, where $z=\exp(i\pi d)$. We have checked that the obtained monodromy matrices satisfy bilinear constraints dictated by twisted Riemann bilinear relations. All examples presented in this paper as well as several additional examples are included in the ancillary files as \textit{Mathematica} notebooks.

Our observation leads to a number of open questions and directions for further research. First, it is natural to ask if the monodromy group for multiloop integrals in $d$ dimensions can be always\footnote{Possibly with the replacement of $\mathbb{Z}$ with another ring of algebraic integers if the discrete symmetries of integrals are taken into account.}  chosen to be a subgroup of $GL(n,\mathbb{Z}[z,1/z])$. A related question is about the form of the similarity transformations which preserve the property of monodromy group of being a subgroup of $GL(n,\mathbb{Z}[z,1/z])$. Of course, the transformation matrices which themselves belong to $GL(n,\mathbb{Z}[z,1/z])$ do preserve this property, but in several examples we have seen that there are also nontrivial transformation matrices from  $GL(n,\mathbb{Q}(z))$ which also preserve this property.

Next, it is interesting  to connect our finding with the results of approach of Ref. \cite{shimada1996picard} via Lee-Pomeransky or Baikov representations. Besides, it is interesting to apply our approach to a problem of finding monodromy group for functions defined via Euler-type integrals\footnote{The analytically regularized multiloop integrals represent a specific example of Euler-type integrals.} \cite{gelfand1990generalized}, in particular, for $A$-hypergeometric functions \cite{gel1989hypergeometric}. Some known results for monodromy groups of generalized hypergeometric functions obtained by different approaches in Ref. \cite{beukers1989monodromy,goto2016monodromy,beukers2016monodromy,goto2022lauricella} support the existence of monodromy representation in $GL(n,\mathbb{Z}[z_1,1/z_1,\ldots,z_m, 1/z_m])$ with $z_k=\exp(2\pi i \alpha_k)$ where $\alpha_k$ are parameters of hypergeometric functions.

\appendix
\section{Eigenvalues of $M_a$ and $\M_a$}\label{sec:appendix}

We prove that all eigenvalues of $\M_a$ are proportional to a rational power of $z$. Thanks to the relation \eqref{eq:local}, it amounts to the proof that all eigenvalues of $M_a$ have the form $c_1 d+c_0$, where $c_1$ is a rational number and $c_0$ is independent of $d$.
First, since the matrix $M(x,d)$ in the original system \eqref{eq:de1} has elements in $\mathbb{Q}(x,d)$, the rational in $x$ transformations which reduce the system to normalized Fuchsian form at $x=a$ are algebraic in $d$ in the worst case. Then the matrix elements as well as the eigenvalues of $M_a$ are also algebraic functions of $d$.

Let $\lambda(d)$ is one of the eigenvalues of $M_a(d)$. Then, since monodromy does not change upon the shift $d\to d+2$ (up to similarity), the matrix $M_a(d)$ should also have an eigenvalue $\lambda_1(d)$, such that $\lambda_1(d)-\lambda(d+2)\in \mathbb{Z}$. Then, on the same basis, the matrix $M_a(d)$ should have an eigenvalue $\lambda_2(d)$, such that $\lambda_2(d)-\lambda_1(d+2)\in \mathbb{Z}$, and, in general, $\lambda_k(d)$, such that $\lambda_k(d)-\lambda(d+2k)\in \mathbb{Z}$. Since there are at most $n$ different eigenvalues, there is a minimal $0<k\leqslant n$ such that
\begin{equation}
	\lambda(d)-\lambda(d+2k)=N\in \mathbb{Z}.\label{eq:lambda}
\end{equation}

Since $\lambda(d)$ is an algebraic function of $d$, it has series expansion at $d=\infty$ in $d^{1/m}$  where $m$ is some positive integer. Let us represent such an expansion as
\begin{equation}
	\lambda(d) =
	\sum_{\nu\in S} d^{\nu}\sum_{j=0}^{\infty} c_{\nu-j} d^{-j}\,,
\end{equation}
where $S$ is a finite set of leading powers, each having distinct fractional part, such that $ c_{\nu}\neq 0$. Then $\lambda(d)-\lambda(d+2k)$ has the expansion
\begin{equation}
	\lambda(d)-\lambda(d+2k) =
	\sum_{\nu\in S} d^{\nu-1}\sum_{j=0}^{\infty} \tilde{c}_{\nu-j} d^{-j},
\end{equation}
where the leading coefficients read $\tilde{c}_{\nu}=-2\nu k c_\nu$. Meanwhile, by Eq. \eqref{eq:lambda}, the right-hand side is a constant integer number $N$. Then it is obvious that the set $S$ can be only $\{1\}$ (if $N\neq 0$), $\{0\}$ (if $N=0$ and $\lambda\neq 0$), or  $\emptyset$ (if $\lambda=0$). In any case, we can rewrite the expansion of $\lambda(d)$ as
\begin{equation}
	\lambda(d) =c_1 d + c_0 + \sum_{j=1}^{\infty} c_{-j} d^{-j}\,,
\end{equation}
Then
\begin{equation}
	N=\lambda(d)-\lambda(d+2k) = -2k c_1+
	\sum_{j=1}^{\infty} \tilde{c}_{-j} d^{-j},
\end{equation}
From this equality it follows that $c_1=-N/(2k)$ is rational and $\tilde{c}_{-j}=0$ for all $j>0$. The latter condition entails that
$c_{-j}=0$ for all $j>0$. Indeed, suppose that $j_0>0$ is a minimal number such that $c_{-j_0}\neq 0$, then $\tilde{c}_{-j_0} = 2j_0 k c_{-j_0}\neq 0$, which is a contradiction.

\bibliographystyle{JHEP}
\bibliography{monodromy.bib}

\end{document}